\documentclass[Afour,sagev,times]{sagej}

\usepackage{moreverb}
\usepackage{url}
\usepackage[dvipdfmx,colorlinks,bookmarksopen,bookmarksnumbered,citecolor=red,urlcolor=red]{hyperref}
\newcommand\BibTeX{{\rmfamily B\kern-.05em \textsc{i\kern-.025em b}\kern-.08em
T\kern-.1667em\lower.7ex\hbox{E}\kern-.125emX}}

\def\volumeyear{2016}

\newtheorem{Def}{Definition}
\usepackage{kbordermatrix}
\usepackage{natbib}
\usepackage{color}
\usepackage{colortbl}
\usepackage{listings}
\begin{document}
\setlength{\abovedisplayskip}{0pt}
\setlength{\belowdisplayskip}{2pt}
\runninghead{Image frequency analysis}

\title{A new radiomics feature: image frequency analysis}

\author{Takuma Usuzaki\affilnum{1}, Kengo  Takahashi\affilnum{2}, Kazuma Umemiya\affilnum{1}}

\affiliation{\affilnum{1}Takeda General Hospital, Aizuwakamatsu, Japan\\
\affilnum{2}Tohoku University Graduate School of Medicine, Sendai, Japan
}

\corrauth{Takuma Usuzaki, MD
Address: Takeda General Hospital
3-27 Yamaga-machi, Aizuwakamatu, Fukushima 965-8585, Japan
Phone: +81-242-27-5511
}

\email{takuma.usuzaki.p6@dc.tohoku.ac.jp}

\begin{abstract}
Radiomics is a promising technology that focuses on improvements of image analysis, using an automated high-throughput extraction of quantitative features. However, the character of lesion is affected by the surrounding tissue. A lesion on medical image should be characterized from the inter-relation between lesion and surrounding tissue as well as property of the lesion itself. The aim of this study is to introduce a new radiomics feature which quantitatively analyze the inter-relation between lesion and surrounding tissue focusing on the value change of rows and columns in a medical image.
\end{abstract}


\maketitle

\section{Introduction}
Radiomics is a method that extracts features from medical images using data-characterization algorithms.\cite{doi:10.1148/radiol.2015151169} Radiomics have been developed based on the concept that medical images contain information reflects underlying pathophysiology.\cite{doi:10.1148/radiol.2015151169} To derive the radiomic feature from medical image, intensity levels, texture heterogeneity patterns and shape of lesion.\cite{LAMBIN2012441} The potential of radiomics has been shown across multiple tumor types, including brain, head and neck, cervix, and lung cancer tumors. Radiomic features extracted from MRI, PET, or CT images, were associated with several clinical outcomes, and hence, potentially provide complementary information for decision support in clinical practice.\cite{vanGriethuysene104} Although radiomics is a promising technology to analyze a medical image, there was a lack of standardization of both feature definitions and image processing.\cite{Stephen2016,Orlhac414,Tixier693} To solve this problem, van Griethuysen et al. constructed \textit{PyRadiomics} which is a flexible open-source platform capable of extracting a large panel of engineered radiomic features from medical images.\cite{vanGriethuysene104} van Griethuysen et al. subdivided radiomics features into 8 classes: First Order Statistics (19 features), Shape-based (3D) (16 features), Shape-based (2D) (10 features), Gray Level Cooccurence Matrix (24 features), Gray Level Run Length Matrix (16 features), Gray Level Size Zone Matrix (16 features), Neighbouring Gray Tone Difference Matrix (5 features), and Gray Level Dependence Matrix (14 features).\cite{vanGriethuysene104} In \textit{PyRadiomics}, these features are evaluated only in an annotated area by mask image: the relation between lesion and surrounding tissue is not evaluated. However, the character of lesion is affected by the surrounding tissue. A lesion on medical image should be characterized from the inter-relation between lesion and surrounding tissue as well as property of the lesion itself.\cite{LAMBIN2012441,Janiszewska2020,https://doi.org/10.1002/jmri.10395} The aim of this study is to introduce a new radiomics feature which quantitatively analyze the inter-relation between lesion and surrounding tissue focusing on the value change of rows and columns in a medical image.

\section{Methods}
\subsection{Definitions of some terms and functions}

$X_{ij}$ denotes the pixel value of $i$-th row and $j$-th column in a $m\times n$ medical image $(1\le i\le m, 1\le j\le n)$. For simplicity, we assume a medical image is a gray-scale image and $X_{ij}$ is integer in the range  $0\le X_{ij}\le 255$.

\begin{Def}[\textbf{Correlation coefficient}]
Let $r_{i,j}$ be the correlation coefficient between $i$-th and $j$-th row. Let $R_{i,j}$ be the correlation coefficient between  $i$-th and $j$-th columns. 
\end{Def}
\begin{align*}
&r_{i,j}=\frac{\sum_{k=1}^n(X_{ik}-\overline{X}_{ik})(X_{jk}-\overline{X}_{jk})}{\sqrt{\sum_{k=1}^n(X_{ik}-\overline{X}_{ik})^2}\sqrt{\sum_{k=1}^n(X_{jk}-\overline{X}_{jk})^2}} \\
&\textrm{where }\overline{X}_{ik}=\frac{1}{n}\sum_{k=1}^nX_{ik}.
\end{align*}

\begin{align*}
&R_{i,j}=\frac{\sum_{k=1}^m(X_{ki}-\overline{X}_{ki})(X_{kj}-\overline{X}_{kj})}{\sqrt{\sum_{k=1}^m(X_{kj}-\overline{X}_{kj})^2}\sqrt{\sum_{k=1}^m(X_{kj}-\overline{X}_{kj})^2}} \\
&\textrm{where }\overline{X}_{kj}=\frac{1}{m}\sum_{k=1}^mX_{kj}.
\end{align*}

\subsection{Explanation of the method}

In this section, we introduce the \textit{Image frequency} and \textit{Image frequency matrix}.

\subsection*{Image frequency}

Image frequency is obtained by applying Fourier analysis to the array of correlation coefficients calculated from adjacent columns or rows. First, we calculate the correlation coefficients between adjacent rows or columns, and obtain array of correlation coefficients. 
\begin{align*}
&[r_{1,2}, r_{2,3}, \cdots, r_{m-2,m-1}, r_{m-1,m}] \\
&[R_{1,2}, R_{2,3}, \cdots, R_{n-2,n-1}, R_{n-1,n}]
\end{align*}
By applying Fourier analysis to these arrays, we obtain frequency of the arrays,
\begin{align*}
&[r_{1,2}, r_{2,3}, \cdots, r_{n-2,n-1}, r_{m-1,m}]  \xrightarrow{\textrm{Fourier analysis}} f\\
&[R_{1,2}, R_{2,3}, \cdots, R_{n-2,n-1}, R_{n-1,n}] \xrightarrow{\textrm{Fourier analysis}} F.
\end{align*}
We call the frequency obtained by these procedure as  \textit{Image frequency}. Note that fourier analysis may result in several frequency. In this case, an additional procedure such as using maximum frequency is needed.

\subsection*{Image frequency matrix for row}

Image frequency matrix for row is obtained by calculating frequency between a $i$-th and $i+M$-th rows. In this procedure, the start row $i_s$ and integer $M (1\le M < m)$ should be determined. The number of the combinations is $m(m-1)$.
\begin{align*}
[r_{i_s,i_s+M}, \cdots, r_{i_s-1,i_s-1+M}] \xrightarrow{\textrm{Fourier analysis}} f_{i_s,N}
\end{align*}
When the indices are over $m$, the indices should be subtracted by $m$. $f_{i_s,N}$ have two indices and can be interpreted as matrix. The diagonal components of $f_{i_s,N}$ are defined as 0.

\subsection*{Image frequency matrix for column}

Similarly, image frequency matrix is obtained by calculating frequency between a $j$-th and $j+M$-th columns. In this procedure, the start column $j_s$ and integer $N (1\le N < n)$ should be determined. The number of the combinations is $n(n-1)$.
\begin{align*}
[R_{j_s,j_s+N}, \cdots, R_{j_s-1,j_s-1+N}] \xrightarrow{\textrm{Fourier analysis}} f_{j_s,N}
\end{align*}
When the indices are over $n$, the indices should be subtracted by $n$. $f_{j_s,N}$ have two indices and can be interpreted as matrix. The diagonal components of $f_{j_s,N}$ are defined as 0.

\setcounter{figure}{0}

\section{Example}

\begin{figure}[htb]
\begin{center}
\includegraphics[width=6cm]{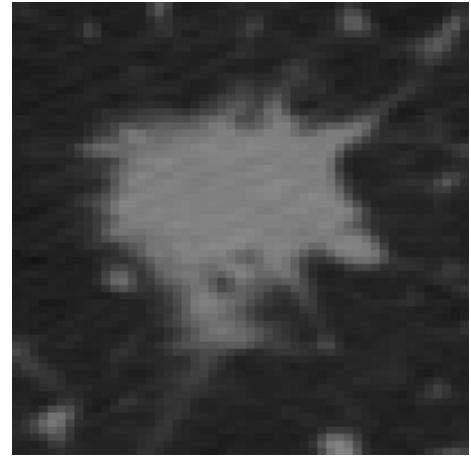}
\caption{The example image for image frequency analysis}
\label{fig:img}
\end{center}
\end{figure}

The images for analysis were obtained from the Lung Image Database Consortium-Image Database Resource Initiative (LIDC/IDRI) constructed by seven academic centers in the US and eight medical imaging companies.\cite{Clark2013} From LIDC/IDRI, an image of a lung nodule which was categorized into highly suspicious as malignant by four radiologists was analyzed as an example. After cropping by annotation data, the image was expanded using the Pillow package in Python with the LNCZOS option set to a $256\times 256$. \textbf{\textsf{Figure}}  \ref{fig:img} shows the example image of lung nodule. 

\subsection*{Image frequency matrix for row}
\begin{figure}[htb]
\begin{center}
\includegraphics[width=9cm]{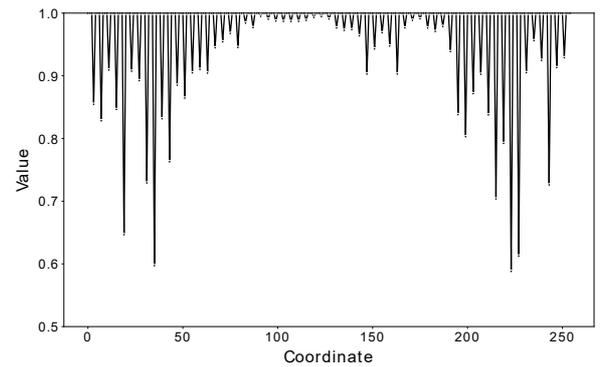}
\caption{Correlation coefficients for rows. Vertical and horizontal axis show value of correlation coefficient and the number of rows, respectively when the image was interpreted as a matrix.}
\label{fig:r_row}
\end{center}
\end{figure}

\begin{figure}[htb]
\begin{center}
\includegraphics[width=9cm]{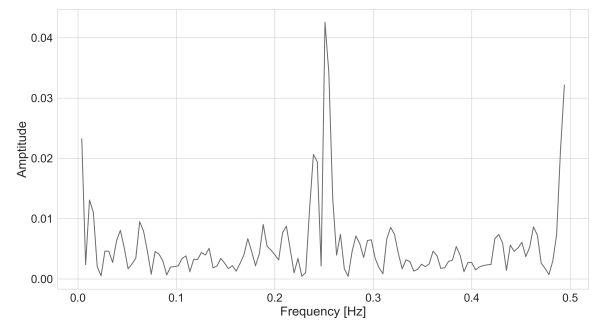}
\caption{The result of Fourier analysis to correlation coefficients of adjacent rows}
\label{fig:f_row}
\end{center}
\end{figure}

\textbf{\textsf{Figure}} \ref{fig:r_row} shows the correlation coefficients for rows. In drawing this figure, we calculate $256-1=255$ correlation coefficients $r_{1,2}, \cdots r_{255,256}$.
\textbf{\textsf{Figure}} \ref{fig:f_row} Applying Fourier analysis to the correlation coefficients, the \textit{Image frequency} 0.255Hz was obtained.

\subsection*{Image frequency matrix for column}
\begin{figure}[htb]
\begin{center}
\includegraphics[width=9cm]{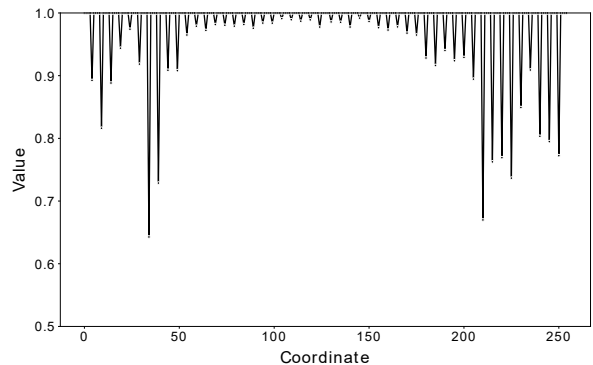}
\caption{Correlation coefficients for rows. Vertical and horizontal axis show value of correlation coefficient and the number of columns, respectively when the image was interpreted as a matrix.}
\label{fig:r_column}
\end{center}
\end{figure}

\begin{figure}[htb]
\begin{center}
\includegraphics[width=9cm]{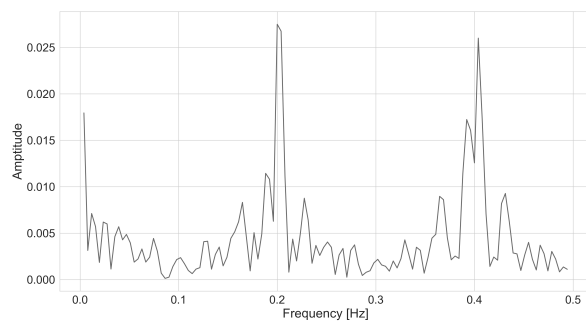}
\caption{The result of Fourier analysis to correlation coefficients of adjacent columns}
\label{fig:f_column}
\end{center}
\end{figure}

\textbf{\textsf{Figure}} \ref{fig:r_column} shows the correlation coefficients for columns. In drawing this figure, we calculate $256-1=255$ correlation coefficients $R_{1,2}, \cdots R_{255,256}$.
\textbf{\textsf{Figure}} \ref{fig:f_column} Applying Fourier analysis to the correlation coefficients, the \textit{Image frequency} 0.204Hz was obtained.

\section{Discussion}

In this short paper, we introduced a new radiomics feature \textit{Image frequency}. \textit{Image frequency} can evaluate the value change of rows or columns in a medical image by applying Fourier analysis to correlation coefficients. In a medical image, benign and malignant lesion have different characteristics.\cite{doi:10.1056/NEJMoa1102873} A benign lesion present morphological feature reflecting its non-invasive property, on the other hand, a malignant lesion present morphological feature reflecting its invasive property. \textit{Image frequency} may capture this difference focusing on value change of rows or columns in a medical image. 

Radiomics feature can reflect biological tumor information, such as cell morphology, molecular, and gene expression.\cite{Lambin2017} Standardized radiomics features in \textit{PyRadiomics}\cite{vanGriethuysene104} have two key concepts. First, morphological features are calculated by information of mask image. Second, change value of pixel values is calculated by several algorithms. Although these radiomics features are powerful tools to evaluate region of interest in a medical image, there is limited number of studies which focuses on value change of rows and columns. In addition to this limitation, previous radiomics study has problems in implementing deep learning model: it is difficult to combine a medical image itself with radiomics feature. This problem is caused by the image array size. In a deep learning model, a medical image is handled as array or matrix, on the other hand, radiomics feature is often scalar value. This differences in array size (matrix vs. scalar) results in a difficulty in implementing deep learning model. Image frequency matrix may solve this problem because image frequency matrix can have the same size as original image in the case that the original image is square. When the original image and image frequency matrix have the same size, these can be merged and learned by deep learning model. From these viewpoints, \textit{Image frequency} and image frequency matrix have new aspects. Further study is needed to investigate the property of \textit{Image frequency} and image frequency matrix.

\section{Conclusion}

We introduced a new radiomics feature called \textit{Image frequency} which quantitatively analyze the inter-relation between lesion and surrounding tissue focusing on the value change of rows and columns in a medical image. Image frequency matrix may enable to construct deep learning model which considers the original medical image and radiomics features at the same time. 

\section{Conflict of Interest}

All authors have no conflict of interest to disclose with respect to this research.

\newpage
\bibliographystyle{SageV}
\bibliography{image_frequency}
\end{document}